\documentclass[12pt]{iopart}

\usepackage{iopams}  

\usepackage{xcolor}
\usepackage{graphicx}
\usepackage{epstopdf}
\usepackage[numbers]{natbib}
\usepackage{hyperref}
\usepackage{comment}
\usepackage{subcaption}

\graphicspath{ {./figs/}, }

\begin{document}

\title[Non-homogeneous, path-dependent anomalous diffusion]{A non-homogeneous, non-stationary and path-dependent Markov anomalous diffusion model}

\author{Nestor R. Barraza$^{1,2,\dagger}$, Gabriel Pena$^{1,2}$,
Juliana Gambini$^{3,4}$ and M. Florencia Carusela$^5$}
\address{$^1$ Departamento de Ciencia y Tecnología, Universidad Nacional de Tres de Febrero, Valentín Gómez 4772, Caseros, B1678ABH, Buenos Aires, Argentina.}
\address{$^2$ Facultad de Ingeniería, Universidad de Buenos Aires, Paseo Colón 850, Ciudad de Buenos Aires, C1063ACV, CABA, Argentina.}
\address{$^3$ LIDEC, Universidad Nacional de Hurlingham, Tte. Origone 151, Villa Tesei, B1688AXC, Buenos Aires, Argentina.}
\address{$^4$ CPSI, Universidad Tecnológica Nacional FRBA, Medrano 951, Ciudad de Buenos Aires, C1179AAQ, CABA, Argentina.}
\address{$^5$ Instituto de Ciencias, Universidad Nacional de General Sarmiento, Juan María Gutiérrez 1150, Los Polvorines, B1613GSX, Buenos Aires, Argentina.}

\ead{$^\dagger$ nbarraza@untref.edu.ar}

\begin{abstract}
    A novel probabilistic framework  for modelling anomalous diffusion is presented. The resulting process is Markovian, non-homogeneous, non-stationary, non-ergodic, and state-dependent. The fundamental law governing this process is driven by two opposing forces: one proportional to the current state, representing the intensity of autocorrelation or contagion, and another inversely proportional to the elapsed time, acting as a damping function. The interplay between these forces determines the diffusion regime, characterized by the ratio of their proportionality coefficients. This framework encompasses various regimes, including subdiffusion, Brownian non-Gaussian, superdiffusion, ballistic, and hyperballistic behaviours. The hyperballistic regime emerges when the correlation force dominates over damping, whereas a balance between these mechanisms results in a ballistic regime, which is also stationary. Crucially, non-stationarity is shown to be necessary for regimes other than ballistic. The model's ability to describe hyperballistic phenomena has been demonstrated in applications such as epidemics, software reliability, and network traffic. Furthermore, deviations from Gaussianity are explored and violations of the Central Limit Theorem are highlighted, supported by theoretical analysis and simulations. It will also be shown that the model exhibits a strong autocorrelation structure due to a position dependent jump probability.
\end{abstract}

\vspace{2pc}

\submitto{\jpa}

\section{Introduction}\label{sec_intro}

Anomalous diffusion has attracted the interest of researchers for decades. This phenomenon has been observed in a wide variety of contexts and in several unrelated fields, including solid-state mechanics, hydrology, biology, and network engineering. Because the classical Gaussian-Markovian solution of the Fokker-Planck equation does not align with experimental results (see, e.g., recent discussions in \cite{Sposini2022, WangBo2012WBdi,C4CP02019G, C4CP03465A}), alternative approaches to describing these types of diffusive phenomena have been introduced over the past few decades. The original approach was the introduction of a generalised diffusion master equation by Scher, Kenkre and Montroll, among others, which enabled the development of non-Markovian diffusion models \cite{Kenkre-Montroll-S, PhysRevB.12.2455}. The next approach was to consider continuous-time random walk (CTRW) models and Lévy walks \cite{ShlesingerDiffusion}; more recent CTRW models can be found in \cite{Metzler_2022_ModellingAnomalous}. Generalisations of Lévy walks have also been recently explored by Albers and Radons \cite{PhysRevLett.120.104501,PhysRevE.105.014113} as diffusion models capable of describing all possible regimes. Other approaches involve processes with diffusivities that are time-dependent, space-dependent, or both, such as those proposed by Cherstvy and Metzler \cite{Cherstvy_2014_SpaceDependentDiffusivities,Cherstvy_2015_TimeSpaceDependentDiffusivities,Cherstvy_2021_TimeDependentDiffusivity}. Superstatistical and diffusing-diffusivity models, where the diffusivity is randomly distributed, have been analysed in \cite{Lemaitre_2023_RandomDiffusivity,Sposini_2018_RandomDiffusivity,Jain2017_DiffusingDiffusivity,Wang_2020_RandomDiffusivity,Wang_2020_UnexpectedCrossover}. One of the most significant advancements was arguably the introduction of models with heavy-tailed distributions of waiting times or displacements, as in the aforementioned Scher, Kenkre and Montroll model or in Lévy flights \cite{Palyulin_2019_LevyWalks}. These models differ significantly from Gaussian processes, exhibiting anomalous properties such as self-similarity or multi-scaling (see e.g. \cite{Stella_2010_AnomalousScaling} for more details on scaling properties). Correlated CTRW models have been proposed in \cite{Tejedor_2010_CorrelatedCTRW_1,Magdziarz_2012_Correlated_CTRW_2}, although not all possible regimes are considered. 

Our proposal is to model the phenomenon of anomalous diffusion using a special family of Markov stochastic processes known as Generalized Polya processes (GPPs). These are a solution of the master equation with state-dependent transitions. This approach is similar to some recent subrecoil laser cooling models, such as the quenched trap model and heterogeneous random walks \cite{PhysRevE.105.064126}. Despite not needing to know the whole process history, the functional form of the jump probability makes the movement path-dependent in the sense that the occurrence of a given displacement influences the future of the trajectory. 

The Polya urn is a discrete model consisting of an urn containing balls of at least two different colours. Each time a ball is extracted, it is returned to the urn along with a given number of additional balls of the same colour. This increases the probability of drawing that colour in the subsequent extraction, thereby modelling a contagion effect. Taking the continuous-time limit leads to the Polya or Polya-Lundberg process \cite{PolyaLundberg}, a stationary Markovian process. GPPs generalise this process by introducing more general time dependencies that allow for non-stationarity. GPPs have been extensively studied in the literature over the past few decades, \cite{konno2010exact,cha_2014,badia_extensions_gpp,fendick_whitt_2021,fendick_whitt_2022}, albeit with few practical applications (see e.g., \cite{le2015recurrent,CSFEpidemiologia, BarrazaRPIC2023,LiGPPReliability}). An interesting property of these processes is the rather general nature of their increments: these are, except for some particular cases, non-homogeneous, non-stationary (and therefore non-ergodic), and strongly correlated, yet they retain the Markov property and exhibit self-similarity. This makes them rather suitable to describe anomalous diffusion behaviour in a broad range of situations, including regimes of subdiffusion, Brownian anomalous diffusion, superdiffusion, ballistic and hyperballistic diffusions. The last two are of particular interest, as they are not often treated but  can be found across different fields. For example, the early stages of an outbreak can be adequately described by an hyperballistic diffusion model, as we have shown in \cite{CSFEpidemiologia} for several case studies corresponding to the 2020 COVID-19 pandemic; the same is true for the early stages of failure detection in modern software development \cite{PenaMorenoBarraza2022}. 

The GPPs are solutions of the discrete Kolmogorov forward equation or non-homogeneous birth equation, also known in statistical physics as the master equation. In a general context, it describes the evolution of a population where individuals can only be born. GPP solutions arise where the birth rate is proportional to the cumulative number of individuals in the population. Heavy-tailed distributions of epochs, correlated increments, and other anomalous diffusion properties emerge naturally in these processes just by choosing proper birth rates. The deviation from Gaussianity, i.e., violation of the Central Limit Theorem (CLT) hypothesis, is shown by directly calculating three critical exponents, namely the Moses, Noah and Joseph scaling exponents. We perform simulations to compute the exponents via time series analysis, as it is usually done in practice. As another property of the GPPs, they excel at modelling contagious phenomena. Contagion is, by definition, the property that occurrence of events increase future jump probabilities, as in the previously described Polya urn model. This means that, although the first event may be difficult to realise, once it occurs, subsequent events will occur with higher probabilities. As previously mentioned, this property is found in several systems across various fields of science and engineering. We have shown that contagion models based on GPPs accurately describe the behaviour of the failure detection process in software development \cite{PenaMorenoBarraza2022} and the spread of epidemics \cite{CSFEpidemiologia,Pena2022MeasuringCS}. More recently, we have studied queuing processes based on GPP with the capability to describe the characteristics of network traffic \cite{BarrazaRPIC2023}.

In addition to introducing a novel application of GPPs, our approach explains the anomalous behaviour through transition probabilities that depend linearly on the current position and inversely on the elapsed time. The former introduces a strong correlation between the increments, whilst the latter induces a non-stationary time dependence via memory-like effects (see \cite{konno2010exact}). The interplay between these two effects determines the resulting diffusion regime. By choosing more general expressions for the time dependence, it becomes possible to model transitions between regimes. This is a path we are currently pursuing, with results to be presented in future works. We aim to highlight the potential applications of this type of processes in contexts of anomalous diffusion, such as heterogeneous and disordered systems.

This paper is organised as follows. In \Sref{sec_gpps}, we briefly describe the theoretical background of GPPs. In \Sref{sec_3pbpm}, we analyse their properties as anomalous diffusion models, described in terms of the scaling exponents. In \Sref{sec_simulaciones}, we present simulations of the proposed process. Finally, in \Sref{sec_conclusiones}, we present our conclusions, summarising the results.

\section{The Generalized Polya Processes as diffusion models}\label{sec_gpps}

The original non-Markovian theory of anomalous diffusion is based on the following equation (see, for example, Equation 11 of \cite{Kenkre-Montroll-S}):
\begin{equation*}
    P(n,t) = \int_0^t \; d\tau \; \psi(t - \tau) \; \sum_{j} \; p(n - j) \; P(j, \tau),
\end{equation*}
where $P(n,t), \; t\geq 0$ and $n,j$ are discrete variables that represent states of the system (e.g. particle positions). This equation denotes the probability of the system being in state $n$ at time $t$, $\psi$ is the waiting time (or pausing time) probability density function (pdf) and $p$ is the jump probability. In this model, $p$ is taken to depend only on the length of the jump (i.e., stationary increments: $p_{n,j}=p(n-j)$) and $\psi$ is taken to depend only on the length of the interval (i.e. time homogeneity: $\psi(\tau, t) = \psi(t - \tau), \; 0<\tau<t$). This equation is a solution of the generalised master equation:
\begin{equation*}
    \frac{\partial}{\partial \; t} \; P(n,t) = \int_0^t d\tau  \; \sum_{j} \; \left[ K_{nj}(t - \tau) P(j,\tau) - K_{jn}(t - \tau) P(n,\tau) \right],
\end{equation*}
where $K_{nj}$ is a kernel function (see \cite{Kenkre-Montroll-S}), and $n, j$ index the states. It is known that the pausing time complementary distribution just satisfies $\Psi(t + s) = \Psi(t) \; \Psi(s)$ for $s, t > 0$, where $\Psi(t) = \int_t^\infty \psi(x)dx$, if and only if the process is Markovian. This condition is satisfied solely by the exponential distribution, so the Scher-Montroll approach was to choose general functional forms of the pausing time different than the exponential in order to obtain non-Markovian solutions. 

Our approach is based on the Kolmogorov forward equation, permitting only transitions to the next site or remaining at the current site. These events have probabilities that depend on both position and time, resulting in non-homogeneous, position-dependent waiting time densities. Let $E_n(t)$ denote the event that ''the particle is at site $n$ by the time $t$''. Following the  postulates of the birth processes (see for example \cite{Feller1}), the probability of jumping in a short time interval $dt$ is proportional to the interval length, while jumps of higher order are not allowed. Thus:
\begin{eqnarray}
    P\left[ E_n(t) \rightarrow E_{n + 1}(t + dt) \right] = \lambda_n(t) \; dt \label{prob_jump} \\
    P\left[ E_n(t) \rightarrow E_{n}(t + dt) \right] = 1 - \lambda_n(t) \; dt
\end{eqnarray}
The proportionality factor $\lambda_n(t)$, usually called birth rate, event rate or intensity function, must be a non-negative and integrable function. Under these conditions, we obtain a diffusion model that is a solution of the following birth equation. This solution gives the conditional discrete probability function, usually named as the probability mass function (pmf) or displacement distribution $p_{k,k+n}(s,t)$ of the particle being at the position $k + n$ by the time $t$, given it was at the position $k$ by the time $s$:
\begin{equation}
    \label{eqp4}
    \frac{\partial}{\partial \; t} p_{k,k+n}(s,t) = -\lambda_{k + n}(t) \;  p_{k,k+n}(s,t) + \lambda_{k + n -1}(t) \;  p_{k,k+n - 1}(s,t) \;\;\; n\geq 1,
\end{equation}
where $k, n \in\mathbb{N}$ and $0 < s < t$. This is a non-stationary master equation (see Equations 1.2 and 3.1 of \cite{konno2010exact}), which corresponds to the convolutionless formalism \cite{hanggi_convolutionless}. This alternative formulation results in anomalous diffusion processes while preserving Markovian dynamics (see also \cite{konno2010exact} for a discussion). This type of master equation with state-dependent transitions, although exhibiting longer-range order,  is also used to model disordered and heterogeneous systems, such as those observed in subrecoil laser cooling (see e.g. \cite{PhysRevE.105.064126,Monthus_2021}).

The resulting waiting time densities take the following form:
\begin{equation}
    \label{generalpsi}
    \psi_n(s, t) = \lambda_{n}(t) \; e^{- \; \int_s^t \; \lambda_n(x) \; dx}.
\end{equation}
The Polya stochastic process emerges as a limiting case of the Polya contagious urn model and is a solution of \Eref{eqp4} with a rate $ \lambda_n(t) = \frac{\beta + \rho \; n}{1 + \rho \; t}$ (see e.g. \cite{Feller1}). Generalized Polya processes arise when more general rates of the following form are considered:
\begin{equation}
    \label{eqp5}
    \lambda_n(t) = (\beta + \gamma \; n) \; \kappa(t),
\end{equation}
for some $\kappa(t) \geq 0$ that depends only on time, and $\beta\geq 0$, $\gamma > 0$. The function $\kappa(t)$, also known as the relaxation function, captures the non-stationary dynamics of the process. In \cite{klugman} it is shown that cases with $\gamma < 0$ are also valid and yield binomial processes; however, these are not included among the GPP and the resulting formulas are completely different. The case $\gamma=0$ reduces to the well-known non-homogeneous Poisson processes, for which the event rate does not depend on the state of the process. The case $\beta=0$ leads to a degenerate process if the initial condition is zero. These cases are not discussed further. In what follows, we always assume that $\gamma$ and $\beta$ are strictly positive. 
The resulting process can indeed model anomalous diffusion with the desired properties, such as self-similarity and non-stationarity, by properly selecting the functional form of the function $\lambda_n(t)$. As stated in \cite{badia_extensions_gpp}, the only conditions imposed on the function $\kappa(t)$ are integrable and non-negative, a very mild setting.

The solution of \Eref{eqp4} for general functional forms of $\kappa(t)$ is Markovian, with non-stationary (and therefore non-ergodic) correlated increments. For the initial conditions $p_{k,k}(s,s)=1$ and $p_{k,k+n}(s,s)=0$, the solution can be expressed recursively as (see \cite{sendova}):

\begin{eqnarray*}
    p_{k,k+n}(s,t) &= \int_s^t \lambda_{k+n-1}(y) p_{k,k+n-1}(s,y) e^{-\int_y^t \lambda_{k+n}(x)dx}dy \qquad n\geq 1\\	
    p_{k,k}(s,t) &= e^{-\int_s^t \lambda_k (x) dx}.
\end{eqnarray*}
The explicit solution of \Eref{eqp4} with the rate given by \Eref{eqp5} is as follows:
\begin{equation}\label{gpp_pmf}
    p_{k,k+n}(s,t) = \frac{\Gamma\left( \frac{\beta}{\gamma} + k + n \right)}{\Gamma\left( \frac{\beta}{\gamma} + k \right)n!} \left[ e^{-\gamma K(s,t)} \right]^{\frac{\beta}{\gamma} + k}\left[ 1 - e^{-\gamma K(s,t)} \right]^n,
\end{equation}
where $K(s,t)=\int_s^t \kappa(x)dx$, and $\Gamma$ denotes the Gamma function. Note that this is a generalised negative binomial distribution (which reduces to the standard form whenever $\frac{\gamma}{\rho} \in \mathbb{Z}$). This and several other distributions involved in GPPs are derived in \cite{cha_2014}. 

Next, we discuss moments. Let $X(t)$ denote the state of the process at time $t$ and assume the initial conditions $s=0$, $k=0$. The mean value function is given by:

\begin{equation}\label{mean_gpp}
    \mu(t) = E[X(t)] = \frac{\beta}{\gamma}\left[ e^{\gamma K(t)} - 1 \right], 
\end{equation}
for $t \geq 0$, where we use $K(t)$ instead of $K(0,t)$ for readability. The autocovariance is given by:

\begin{equation}\label{cov_gpp}
    Cov[X(s),X(t)] = \frac{\beta}{\gamma}  e^{\gamma K(t)} \left[ e^{\gamma K(s)} - 1 \right].
\end{equation}
The variance can thus be obtained straightforwardly as:

\begin{equation}
    \label{variance}
    Var[X(t)] =  Cov[X(t),X(t)] = \frac{\beta}{\gamma}  e^{\gamma K(t)} \left[ e^{\gamma K(t)} - 1 \right].
\end{equation}
Refer to \cite{fendick_whitt_2022} for the derivation of these formulae.

The autocorrelation function, defined as follows, is essential for measuring the decay of long-range order:
\begin{equation}
    \label{autocorrelation}
    R[X(s),X(t)]=\frac{Cov[X(s),X(t)]}{\sqrt{Var[X(s)] \cdot Var[X(t)]}}.
\end{equation}
It is customary to fix $s$ and compute the autocorrelation as a function of $t$ for a wide range of values $t \gg s$. Since $\kappa(t)$ is integrable, it follows that $K(t)$ is finite for all $t < \infty$, ensuring that both the first and second order moments always exist and are finite. Note again that if we set $r=\beta/\gamma$ and $p=e^{-\gamma K(t)}$ then all of these expressions match the known formulae for the negative binomial moments, which remain valid even when $\gamma/\beta$ is not an integer. 

To assess the tail behaviour we also consider the excess kurtosis $EK$, given by:
\begin{equation*}
    EK = \frac{\gamma}{\beta}\left[6 + \frac{e^{-\gamma K(t)}}{e^{\gamma K(t)} - 1} \right],
\end{equation*}
which is strictly positive. This means that the tails are always heavier than normal, with a weight that grows proportionally with $\gamma/\beta$.

The distribution of increments (displacements) is negative binomial (see \cite{cha_2014}):
\begin{eqnarray}
    \label{pmf_inc}
    &P[\delta(s,t)=n] := P[X(t) - X(s) = n] = \nonumber\\
    &\frac{\Gamma\left( \frac{\beta}{\gamma} + n \right)}{\Gamma\left( \frac{\beta}{\gamma}\right)n!}\left[ \frac{1 - e^{-\gamma K(s,t)}}{1 - e^{-\gamma K(s,t)} + e^{-\gamma K(t)}} \right]^n \left[ \frac{e^{-\gamma K(t)}}{1 - e^{-\gamma K(s,t)} + e^{-\gamma K(t)}} \right]^\frac{\beta}{\gamma}.
\end{eqnarray}
It is evident that its moments have the same scaling as those of the marginal distribution (Equations \ref{mean_gpp} and \ref{variance}).

As explained earlier, the original approach to modelling anomalous diffusion was to generalise the master equation through convolution with a kernel function. This function allows the equation to take into account the entire history of the process, creating a complex dependence structure. In that context, non-Gaussianity is caused by the non-Markovianity introduced by the kernel. It is clear that this is not a necessary condition, as the GPPs successfully describe anomalous diffusion while retaining the Markov property. In this approach, the deviation from Gaussianity is explained through two mechanisms: non-stationarity and non-independence of increments. All GPPs are non-stationary except for the Polya process \cite{fendick_whitt_gaussmarkov}, a particular case in which the anomaly is fully explained by the correlations between increments. In this instance, the non-stationary induced by $\kappa(t)$ is equally compensated by the strong correlation effect. This suggests that, since the probability of a particle to jump from one position to the next depends on both time and position, it is reasonable to expect the increments to be strongly correlated. In the GPP family, these dependencies are introduced through the function $\kappa(t)$ and the $\gamma$ parameter respectively. Dependence on the state creates a tendency of the random variable to increase in an accelerated fashion, which can be broadly interpreted as a contagion effect, as it was originally proposed by Polya \cite{PolyaLundberg}. 

\section{The three-parameter BPM proposed model}\label{sec_3pbpm}

To highlight the results obtained from our model, we study the properties of the following rate:

\begin{equation}
    \label{rate}
    \lambda_n(t) = \frac{\beta + \gamma \; n}{1 + \rho \; t},
\end{equation}
where $\rho > 0$, i.e., we take $\kappa(t) = (1 + \rho t)^{-1}$; this is the same as the well-known Omori formula from seismology (see \cite{app12199965,Guglielmi2016}). Note that it depends on three parameters, and is therefore more general than the similar rate considered in \cite{CSFEpidemiologia,PenaMorenoBarraza2022,BarrazaRPIC2023}. We shall call this the three-parameter BPM model (3p-BPM for short, named after the original authors Barraza, Pena and Moreno). 
As will be discussed next, the parameters $\gamma$ and $\rho$ play a crucial role in the process, with their ratio determining the diffusion regime. While $\gamma$ quantifies the extent of the correlation between increments, acting as a forward effect, $\rho$ measures the intensity of the damping function over time, serving in this case as a backward force.

The waiting time thus takes the form:
\begin{equation*}
    \psi_n(s, t) = \frac{\beta + \gamma n}{1 + \rho t} \left( \frac{1 + \rho s}{1 + \rho t} \right)^{\frac{\beta + \gamma n}{\rho}},     
\end{equation*}
which for large $t$ behaves as
$$\psi_n(s,t) \propto t^{-(1 + \alpha_n)},$$ 
where $\alpha_n > 0$. This implies that the 3p-BPM has a power-law waiting time, and is therefore heavy-tailed and self-similar, which is also accomplished by the pdf of the epochs (see Theorem 5 of \cite{BarrazaRPIC2023}). Scher and Montroll considered a heavy tailed waiting time, namely $\psi(t) = 4W e^{-Wt} i^2 \; \mathrm{erfc}(\sqrt{Wt})$ (see Equation 21 of \cite{PhysRevB.12.2455}), which decays as $t^{-3/2}$, and aimed for an extension to general power-law tails. Our model provides such generalised framework.

In summary, the main innovations of our proposal are:
\begin{itemize}
    \item Markovianity in all diffusion regimes. Although this property is shared with some generalized Lévy walks, most anomalous diffusion models are inherently non-Markovian.
    \item Non-homogeneity of space, in the sense that the waiting time depends on the position.
    \item The strong correlation structure present within all time scales.
    \item Except for the ballistic case, non-stationarity in time. 
\end{itemize}     

\subsection{Time scaling and diffusion regimes}

In this section, we show that our model allows for several distinct diffusion regimes. By  Equations \ref{mean_gpp} and \ref{variance}, with the proposed $\kappa(t)$, it can be seen that for large $t$, the moments of orders 1 and 2 scale with exponents that are proportional to the order of the moment, by a factor of $\gamma/\rho$. This type of scaling is a typical property of self-similar processes \cite{VollmerAnomalousDiffusion2021}. Another remarkable property of the 3p-BPM process is that its autocorrelation function scales as a constant, as it is easily seen from Equations \ref{cov_gpp} and \ref{variance}. This means that, when $t\to\infty$, the autocorrelation tends to an asymptotic value greater than zero (see the Supplementary Material for detailed calculations). This reflects the dependence structure of the process: the influence of early conditions does not dissipate over time. This fact, known as the absence of asymptotic loss of memory (ALOM), was already pointed out in \cite{fendick_whitt_2022}. The scaling matches exactly that of the Scher-Montroll model when setting $\alpha = \gamma/\rho$ ($\alpha$ being the parameter defined in Equation 31 of \cite{PhysRevB.12.2455}); however, it must be remarked that the models are \textit{not} equivalent. In our model, the interarrival time $\psi(t)$ depends on the current position, where in the Scher-Montroll model it depends solely on time.

\begin{figure}[!ht]
   \centering
   \includegraphics[width=0.8\textwidth]{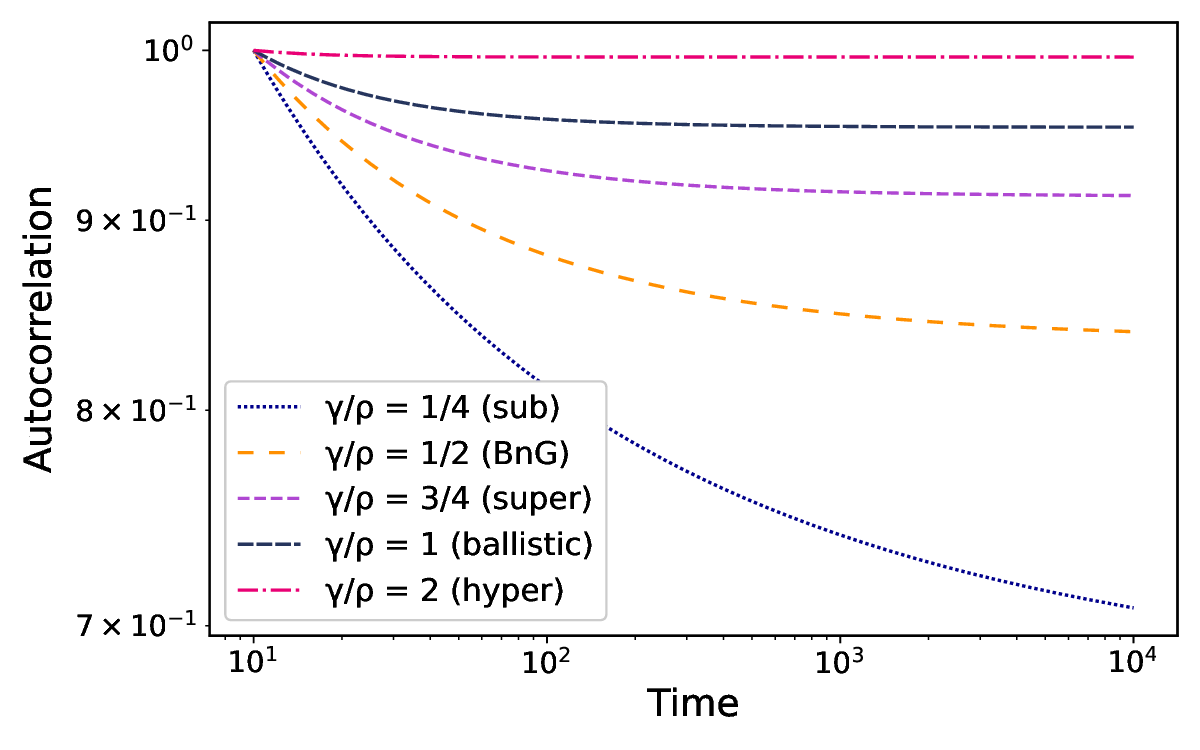}
   \caption{Autocorrelation function for the 3p-BPM process for different values of $\gamma/\rho$ in log-scale. }
   \label{graf_acorr_gpp_log}
\end{figure}

\Fref{graf_acorr_gpp_log} depicts the autocorrelation function in logarithmic scale for different choices of $\gamma/\rho$. It can be observed that the autocorrelation does not decay to zero, but instead converges to a constant value when $t\to\infty$, as previously discussed. Moreover, the asymptotic limit depends on the $\gamma/\rho$ parameter (as can be observed from \Eref{autocorrelation} for large $t$; refer to the Supplementary Material). It approximates 1 as $\gamma/\rho$ becomes larger (i.e. hyperballistic regimes) and 0 as $\gamma/\rho$ becomes smaller (i.e. subdiffusion). This is consistent with the contagion effect: on the one hand, it will be less dominant as the quotient $\gamma/\rho$ becomes smaller, and the process will resemble a NHPP. On the other hand, the contagion effect will become more prominent as $\gamma/\rho$ increases, and the long-range dependence will be greater.

As previously stated, a key property of the 3p-BPM process is its capability to model any diffusion regime. Since the MSE scales as $t^{2\frac{\gamma}{\rho}}$, the Hurst exponent is $H=\frac{\gamma}{\rho}$, therefore the process behaves as:  
\begin{itemize}
    \item Subdiffusion: $0 < \gamma/\rho < 1/2$.
    \item Brownian non-Gaussian (BnG) diffusion: $\gamma/\rho = 1/2$.
    \item Superdiffusion: $1/2 < \gamma/\rho < 1$.
    \item Ballistic diffusion: $\gamma/\rho = 1$.
    \item Hyperballistic diffusion: $\gamma/\rho > 1$.
\end{itemize}

\begin{figure}[!htbp]
    \centering
    \includegraphics[width=0.8\textwidth]{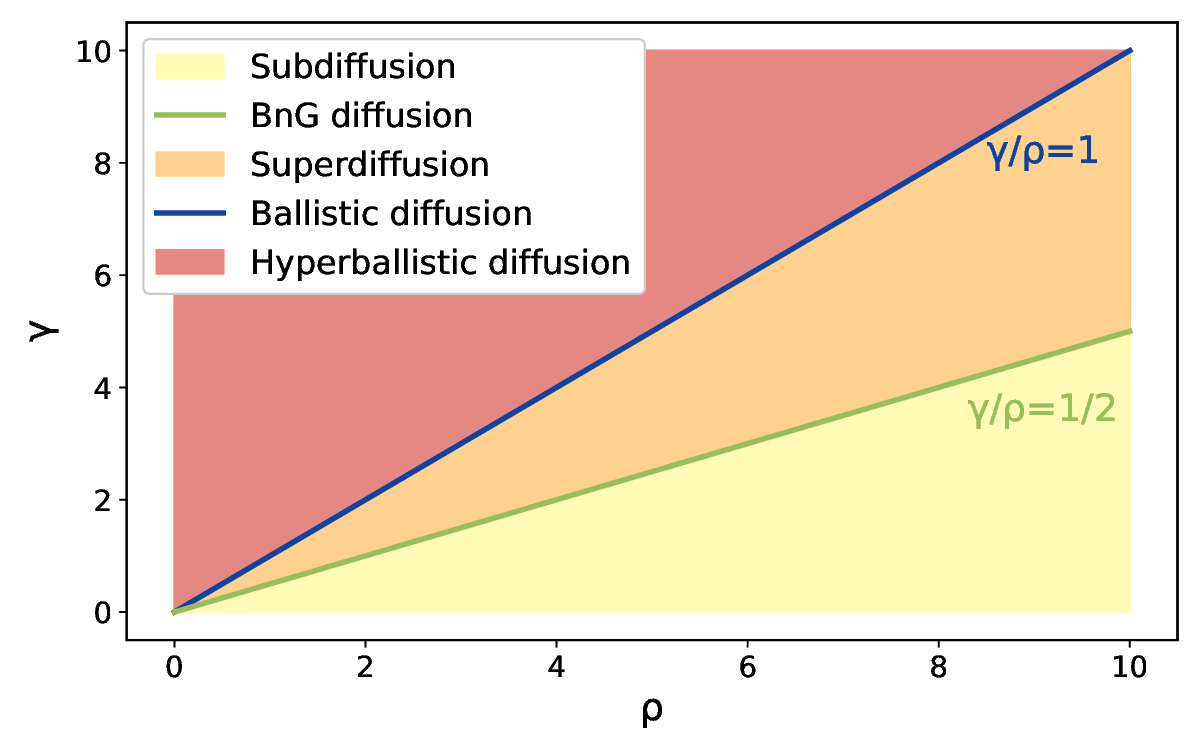}
    \caption{All different diffusion regimes depicted as a phase diagram plot of $\gamma$ vs $\rho$.}
    \label{fig_df_difusiones}
\end{figure}

\begin{figure}[!ht]
    \centering
    \begin{subfigure}{0.95\textwidth}
        \centering
        \includegraphics[width=\textwidth]{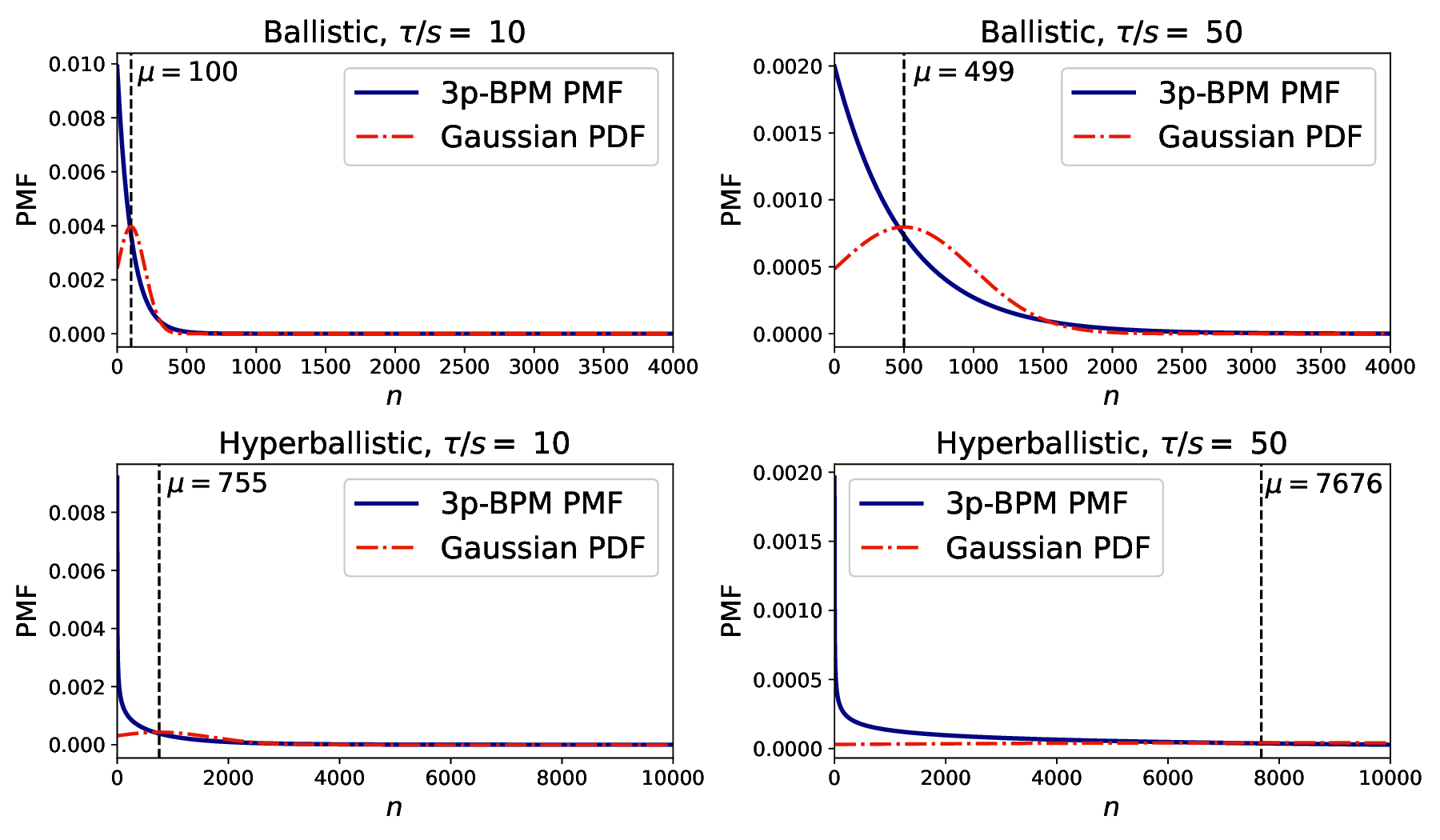}
        \caption{Linear scale.}
        \label{subfig_increments_linear}
    \end{subfigure}
    \\
    \begin{subfigure}{0.95\textwidth}
        \centering
        \includegraphics[width=\textwidth]{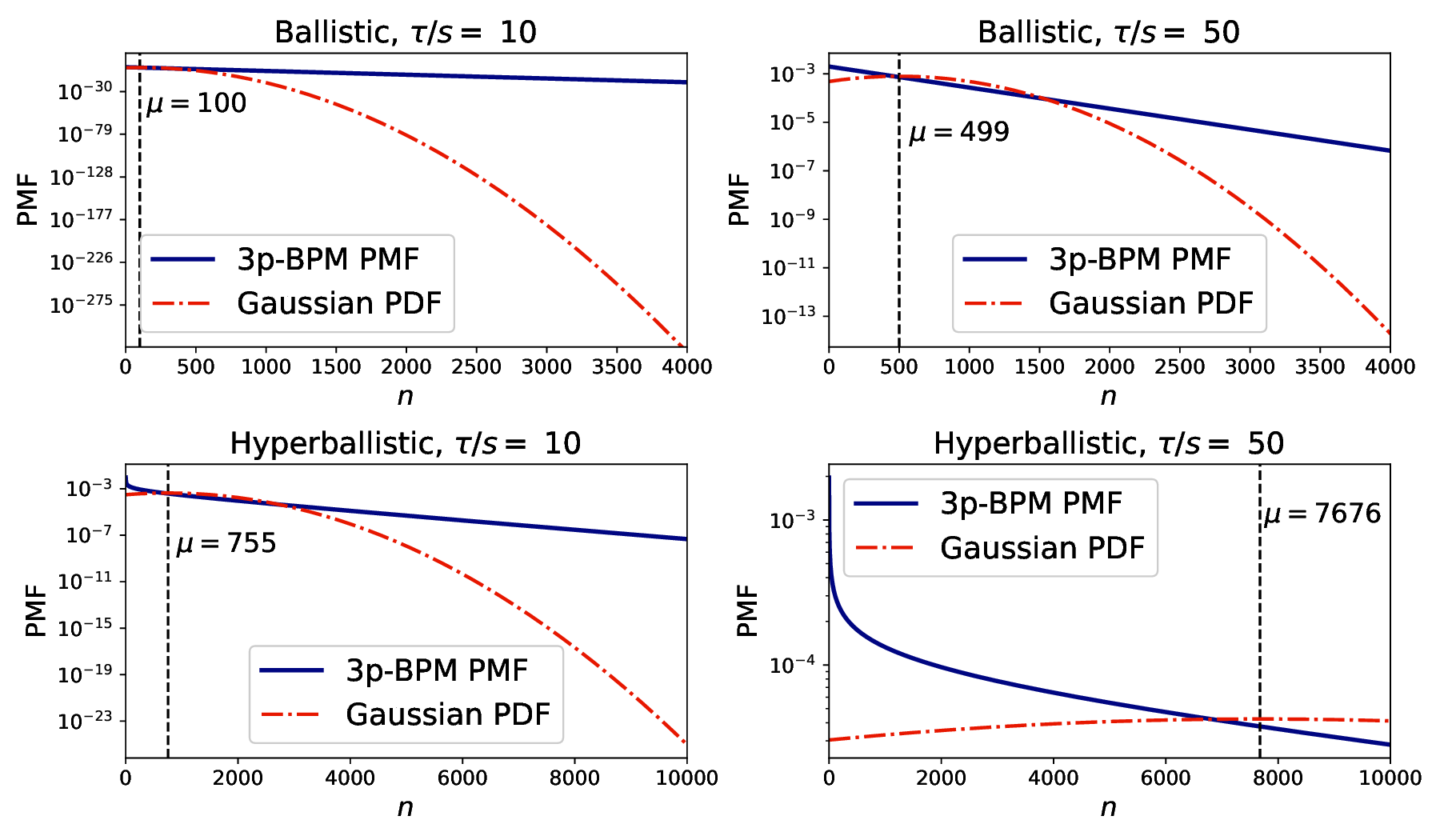}
        \caption{Log-linear scale.}
        \label{subfig_increments_log}
    \end{subfigure}
    \caption{Displacement distribution (pmf) $X(s + \tau) - X(s)$ of the 3p-BPM model, for ballistic and hyperballistic cases and for different time scales $\tau/s$. The dash-dotted line is the  Gaussian density with the same mean and variance. The common mean is marked by the vertical dashed line. The excess kurtosis attains a value of 6 for the ballistic and 9 for the hyperballistic case, being independent of $\tau/s$. For the ballistic we used $\beta = \gamma = \rho = 1$ and for the hyperballistic we used $\beta = \rho = 1$ and $\gamma = 1.5$.}
    \label{fig_increments}
\end{figure}

This shows that the 3p-BPM model can be used to describe any kind of diffusion regime, regardless of its speed; a graphical summary is shown as a phase diagram in \Fref{fig_df_difusiones}, where we depict the $(\rho, \gamma)$ parameter space indicating the regions that correspond to each possible regime. Note that the Hurst parameter $H=\gamma/\rho$ can have any positive value, including $1/2$. It must be remarked that the distribution of the process is never Gaussian. This behaviour is different from that of the models proposed in \cite{bng_diffusion,WangBo2012WBdi,Doerries_2023_EmergentAnomalous,Wang_2020_UnexpectedCrossover,Nampoothiri_2022_BrownianNonGaussian}, where the process is always Fickian, but only not Gaussian within certain time scales, below some crossover time. Our model, in contrast, is anomalous at every time scale. This is evident in \Fref{subfig_increments_linear}, where the pmf of the increments (\Eref{pmf_inc}) is plotted for different time scales and compared to a Gaussian distribution displayed in log-linear scale in \Fref{subfig_increments_log}; in this case, the logarithmic scale makes easy to grasp how slow the tail decay of our model is in contrast to the Gaussian. Nonetheless, it must be remarked that despite the tails being heavy, they are not heavy enough for the variance to diverge.

Recalling that the birth rate $\lambda_n(t)$ serves as the proportionality constant of the infinitesimal transition probability (see, for example, \cite{Feller1} or \cite{cha_2014}), and it is proportional to the current position through the parameter $\gamma$ and inversely proportional to time through the parameter $\rho$. Thus the ratio $\gamma/\rho$ can be interpreted as a competition between two opposite forces of acceleration (represented by $\gamma$) and deceleration (represented by $\rho$), both of which influence the resulting event rate. This observation was previously noted in \cite{CSFEpidemiologia}. This interpretation gains more relevance in the context of anomalous diffusion, since the ratio $\gamma/\rho$ is the Hurst parameter. Recall that setting $\rho = \gamma$ in \Eref{rate} reduces to that of the classical Polya process, a particular case of our model. Since the Polya process is the classical description of a stochastic contagious effect, it is reasonable that its diffusion regime aligns with that of ballistic diffusion. Our model is then able to describe all possible diffusion regimes, even those that exhibiting hyperballistic characteristics (see, for example, the simulation of queues in \cite{BarrazaRPIC2023}). 

The random phenomenon governing the movement is as follows. The probability of jumping, determined by the rate $\lambda_n(t)$ (\Eref{prob_jump}), may increase, decrease, or remain unchanged, depending on the relationship between the correlation parameter $\gamma$ and the damping parameter  $\rho$. The hyperballistic regime arises when this probability increases due to positive feedback. As a result, jumps occur more frequently over time. Conversely, in the subdiffusive regime, time damping dominates state dependence, leading to a decreasing jump probability. Ballistic diffusion occurs when the probability of jumping remains unchanged; in this case, the correlation and damping forces are in equilibrium, resulting in a stationary process. Through successful applications of our model, we have discovered that this phenomenon actually occurs in epidemics \cite{CSFEpidemiologia} and software reliability \cite{PenaMorenoBarraza2022}, and it is postulated to appear, though not yet proven, in network traffic \cite{BarrazaRPIC2023}. Another potential field of application is astrophysics (see e.g. \cite{PerriAmatoZimbardo,Zimbardo}).

Several different approaches to describe random hyperballistic motion (also known as superballistic or hyperdiffusion) are based on generalised Lévy walks \cite{PhysRevLett.120.104501,PhysRevE.105.014113} and the generalised Langevin equation \cite{Hanggi}. The key distinctions of our model are its position-dependent waiting time and the ability to explain the phenomenon as a correlation force overcoming a damping or relaxation action, leading to a non-stationary process. Furthermore, while the hyperballistic regime in our model is always non-stationary, it remains Markovian unlike other models, such as that described in \cite{HyperballisticTransport}.

\subsection{Non-Gaussianity and the CLT}

As described in Appendix A of \cite{VilkAnomalousDiffusion2022}, the properties of the increments determine whether the process -understood as an aggregation of its increments- is or is not Gaussian, i.e., whether it obeys the CLT or not. The following conditions are required for the CLT to hold:

\begin{enumerate}
    \item Increments are stationary.
    \item Increments have finite variance.
    \item Increments are uncorrelated \footnote{This can be relaxed to a weaker statement; however, it is usually stated this way in the literature on anomalous diffusion. Nonetheless, since in our model increments are strongly correlated, violation of this condition is guaranteed.}.
\end{enumerate}

If any of these conditions is not satisfied, then one cannot use the CLT to ensure Gaussianity of the aggregated process. Identifying which of these conditions is violated in each case of application may provide insight on the origins of the anomalous diffusion. Violation of condition (i) is called Moses effect, violation of condition (ii) is called Noah effect and violation of condition (iii) is called Joseph effect; anomalous diffusion can exhibit any one or more of these effects. These violations can be quantified via the three exponents $M$, $L$ and $J$, which relate to the Hurst exponent \cite{AghionMosesNoahJoseph2021}. These can be empirically estimated using time series methods, which is tremendously useful when one does not have a tractable model. In \Sref{sec_simulaciones} we pursue both approaches: first deriving the exponents analytically and then estimating them through simulations, which show good agreement. 

Since a GPP is never Gaussian, a violation of conditions of the CLT necessarily has to be present within the increments. This implies that completely normal (Gaussian) diffusions do not arise in the context of GPPs. One can immediately discard violation of condition (ii): distribution of the increments is known in the model and has finite variance for any choice of parameters. Thus a GPP never exhibits Noah effect. Since the Polya process is the only GPP with stationary increments, it is the only one that does not exhibit Moses effect. Condition (iii) is never satisfied: increments are always correlated, as can be verified by taking into account the joint distribution of any number of increments. Therefore the Joseph effect is always present. In particular, for the Polya process we know that conditions (i) and (ii) hold, so the anomalous diffusion is completely dominated by Joseph effect. The complete analytical deduction of the exponents for a 3p-BPM model is presented in the Supplementary Material.

\section{Estimation of the exponents} \label{sec_simulaciones}

In this section we explore the violations of the three CLT conditions via the standard approach using simulations. In what follows, let $X(t)$ denote the state of the process at time $t$ and $\delta X_{j,\Delta} = X(j\Delta) - X[(j-1)\Delta]$ denote the increment within a time gap $\Delta$, for $j=2,3, \dots$. 

\subsection{Moses}

The coefficient $M$, called Moses exponent, measures deviation from stationarity within the increments. It is defined as the scaling exponent of the absolute velocity time average (see \cite{AghionMosesNoahJoseph2021,VilkAnomalousDiffusion2022}):
\begin{equation}\label{moses_ts}
    \left\langle \overline{|v(t)|} \right\rangle = 
 \left\langle \frac{\Delta}{t}\sum_{j=1}^{t/\Delta} \frac{|\delta X_{j,\Delta}|} {\Delta} \right\rangle \propto t^{M - 1/2}
\end{equation}
over an ensemble of time series. In this equation, $\Delta$ is a time lag over which the increments are computed, the overline represents time averaging and the brackets $\left\langle \cdot \right\rangle$ represent ensemble averaging, which is necessary to reduce the noise introduced by using a single sample path. The time lag $\Delta$ must be greater than or equal to the sampling period of the time vector but not nearly as large as the observed times ($\Delta \ll t_{min}$, $t_{min}$ being the least observed time). 

The increments are stationary when $M=1/2$. In Section 1 of the Supplementary Material we show that in a 3p-BPM model $M=\frac{\gamma}{\rho} - \frac{1}{2}$, implying that Moses effect always happens when $\gamma\neq\rho$, which is consistent with the fact that the Polya process $\left(\frac{\gamma}{\rho} = 1\right)$ is the only GPP with stationary increments. Notice that, as this implies, stationarity is not compatible with normal (BnG) diffusion, which occurs when $\frac{\gamma}{\rho} = 1/2$.

\subsection{Noah}

The coefficient $L$, called latent exponent, quantifies the likelihood of extreme events. It is defined as the scaling exponent of the square velocity time average (see \cite{AghionMosesNoahJoseph2021,VilkAnomalousDiffusion2022}):
\begin{equation}\label{noah_ts}
    \left\langle \overline{v^2(t)} \right\rangle = \left\langle \frac{\Delta}{t}\sum_{j=1}^{t/\Delta} \left( \frac{\delta X_{j,\Delta}}{\Delta} \right)^2 \right\rangle \propto t^{2L + 2M - 2},
\end{equation}
with the restriction $1/2 \leq L < 1$. 

In Gaussian diffusion, the variance (as well as the MSD) grows linearly in time; as it is shown in Appendix A of \cite{VilkAnomalousDiffusion2022}, values of $L$ larger than $1/2$ are an indicator of the presence of extreme events, where ``extreme'' means far from Gaussian. This happens due to heavy-tailed increments having infinite variances \cite{chen_anomalous}. When $L=1/2$ there is no Noah effect and the mean square velocity scales as $t^{2M-1}$; this is exactly what happens in GPPs, no matter the choice of the parameters or the function $\kappa(t)$. A detailed proof by direct computation of \Eref{noah_ts} is presented in Section 2 of the Supplementary Material.

\subsection{Joseph}

The coefficient $J$, called Joseph exponent, measures the degree of correlation between increments. It is defined as the scaling exponent of the velocity autocorrelation function with respect to the time gap $\Delta$:
\begin{equation}\label{joseph}
    \int_0^\Delta  \frac{\left\langle v(t) \cdot v(t + \tau) \right\rangle }{\left\langle v^2(t) \right\rangle} d\tau \propto \Delta^{2J - 1},
\end{equation}
with the restriction $0 < J \leq 1$. Computational considerations often lead to estimating the Joseph exponent through the ensemble-time averaged mean square displacement (ETAMSD) rather than its definition, as follows:
\begin{equation}\label{etamsd}
    \left\langle \overline{\delta^2(T, \Delta)} \right\rangle = \left\langle \frac{1}{N - m + 1} \sum_{k=0}^{N-m} \left(X(kh + \Delta) - X(kh) \right)^2  \right\rangle \propto \Delta^{2J},
\end{equation} 
where $T$ is the total observation time, $h$ is the sampling period of the time series and $T = Nh$, $\Delta = mh$ (see \cite{vitali_anomalous_diffusion}). For a proof of the validity of \Eref{etamsd} to estimate $J$, consult Appendix B of \cite{VilkAnomalousDiffusion2022} or Appendix B of \cite{AghionMosesNoahJoseph2021}.

The increments are uncorrelated when $J=1/2$; this is the case of regular Gaussian diffusion and NHPPs. In a GPP the increments are always correlated, which is reasonable since the event rate $\lambda_n(t)$ depends on the current state: the Joseph effect is always present. In fact, in the 3p-BPM model $J=1$ for any choice of parameters, as demonstrated in Section 3 of the Supplementary Material. 

\subsection{Fundamental relation}

The main parameter that characterizes anomalous diffusion is the Hurst parameter, which measures the scaling of the MSD:
\begin{equation}
    \label{hurst}
    \left\langle X^2(t) \right\rangle \propto t^{2H}.
\end{equation}

The Moses, Noah and Joseph exponents can be viewed as a decomposition of the Hurst parameter, each of them quantifying a different aspect of the anomalous diffusion. The following summation relation holds (see \cite{AghionMosesNoahJoseph2021}):
\begin{equation}\label{relation_H}
    H = M + L + J - 1.
\end{equation}

\subsection{Simulations}

We performed several simulation experiments with the 3p-BPM process in order to measure all four exponents and discuss all possible diffusion regimes, confirming our theoretical predictions. Numerical results are shown in \Tref{tab_simulaciones}, which summarises the estimated values of each exponents for different values of $\gamma/\rho$ covering all possible diffusion regimes. For details regarding the simulation parameters, refer to the Supplementary Material. All simulations, calculations and figures shown here were executed using Python; the source code is available at \cite{penagithub-nhbpgen}.

\begin{table}[!ht]
    \renewcommand{\arraystretch}{1.5}
    \caption{Estimation of the four exponents.}
    \label{tab_simulaciones}
    \centering
    \begin{tabular}{|c|c|c|c|c|}
        \hline
        $\gamma/\rho$ & Moses (M) & Noah (L) & Joseph (J) & Hurst (H) \\
        \hline\hline
        $1/4$ & $-0.151$ & $0.674$ & $0.612$ & $0.263$ \\
        \hline
        $1/2$ & $0.046$ & $0.562$ & $0.818$ & $0.505$ \\
        \hline
        $3/4$ & $0.268$ & $0.510$ & $0.967$ & $0.753$ \\ 
        \hline
        $1$ & $0.499$ & $0.500$ & $0.995$ & $0.999$ \\ 
        \hline
        $5/4$ & $0.741$ & $0.506$ & $1.000$ & $1.241$ \\
        \hline
        $3/2$ & $0.993$ & $0.517$ & $0.999$ & $1.477$ 
        \\
        \hline
        $2$ & $1.539$ & $0.548$ & $0.980$ & $1.904$ 
        \\
        \hline
    \end{tabular}
\end{table}

Measurements obtained from the simulation are in agreement with theoretical predictions; values are shown in \Tref{tab_simulaciones}. First, note that for each experiment, the exponents follow (roughly) the summation relation (\Eref{relation_H}). The Hurst exponent is always around $\gamma/\rho$, as we know.

Observe that the Noah exponent $L$ is always near $1/2$, as expected since all variances are finite (the only exception being the subdiffusive case, due to computational limitations). The Joseph exponent is always close to $1$, as expected, save random fluctuations introduced by simulation limitations. It can be seen that $M$ is a linear function of $\gamma/\rho$ with slope $1$ and values about $1/2$ below $H$, so the Moses exponent behaves as expected. The linear shape of $M$ shows that the degree of non-stationarity increases when $\gamma/\rho \to \infty$ and also when $\gamma/\rho\to 0$. Non-stationarity can then be seen as a measure of deviation from the Polya regime (among the 3p-BPM family).

As previously noted, non-negligible deviations from model predictions were observed in the subdiffusion and, to a lesser extent, in the BnG diffusion simulations. More specifically, the slow pace of the process in those cases implies that, as $s$ and $t$ grow larger, most increments $\delta(s,t)$ will tend to be small or even zero, necessitating very large time ranges to detect scaling effects. This becomes more pronounced in the case of the Noah effect, since squaring the increments amplifies the measurement errors. We confirmed this by performing larger simulations. These yielded lower, closer to $1/2$ values of $L$ and higher, closer to $1$ values of the $J$, as expected. However, achieving the desired accuracy proved challenging in these cases. In fact, in the subdiffusive case, increasing the length of the simulation in one order of magnitude after $T=10^5$ barely impacts on the second decimal. Superdiffusive and faster regimes do not exhibit the same sensitivity to simulation length: even with simulation times on the order of $T=10^3$ or less, the results are in good agreement with model predictions, which is reasonable due to the contagious nature of the processes.

\begin{figure}[!ht]
    \centering    
    \caption{Ensemble-time averages for a 3p-BPM process with $\gamma/\rho = 3/4$ (superdiffusion).}
    \begin{subfigure}{0.48\textwidth}
        \centering
        \includegraphics[width=\textwidth]{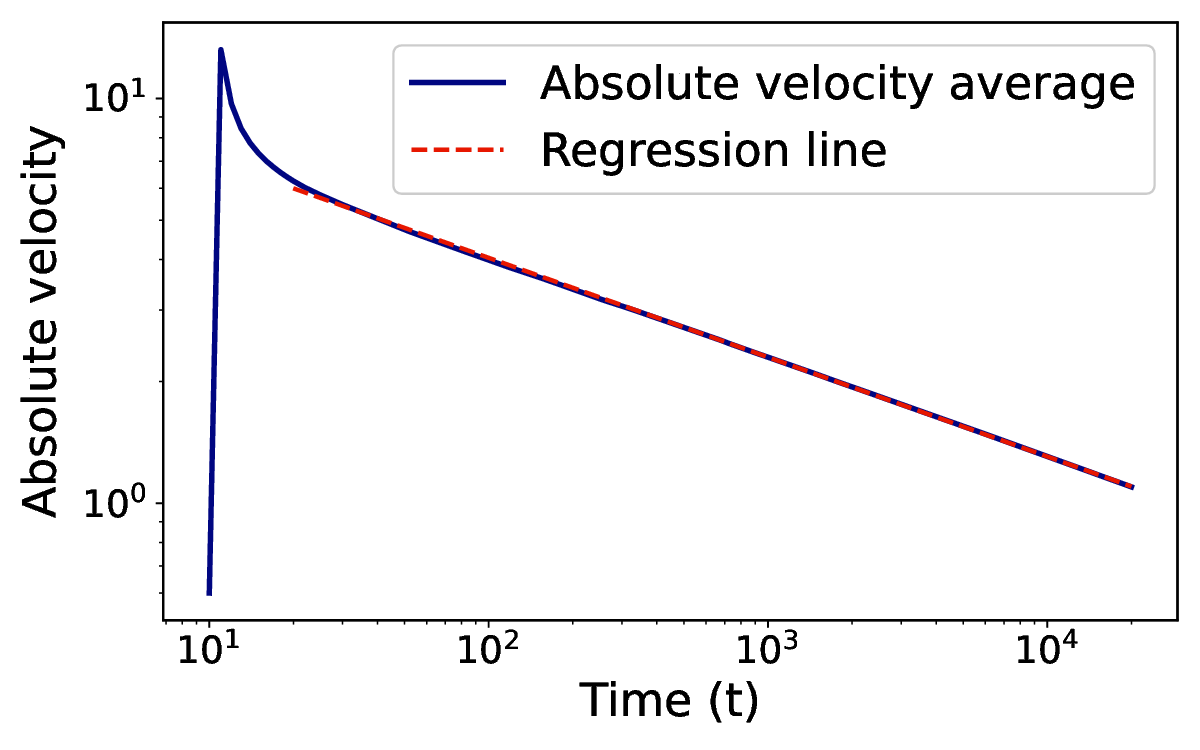}
        \caption{Absolute velocity time-average. Slope is $-0.245$, fit with $R^2 = 0.994$.}
        \label{graf_moses}
    \end{subfigure}
    \hfill
    \begin{subfigure}{0.48\textwidth}
        \centering
        \includegraphics[width=\textwidth]{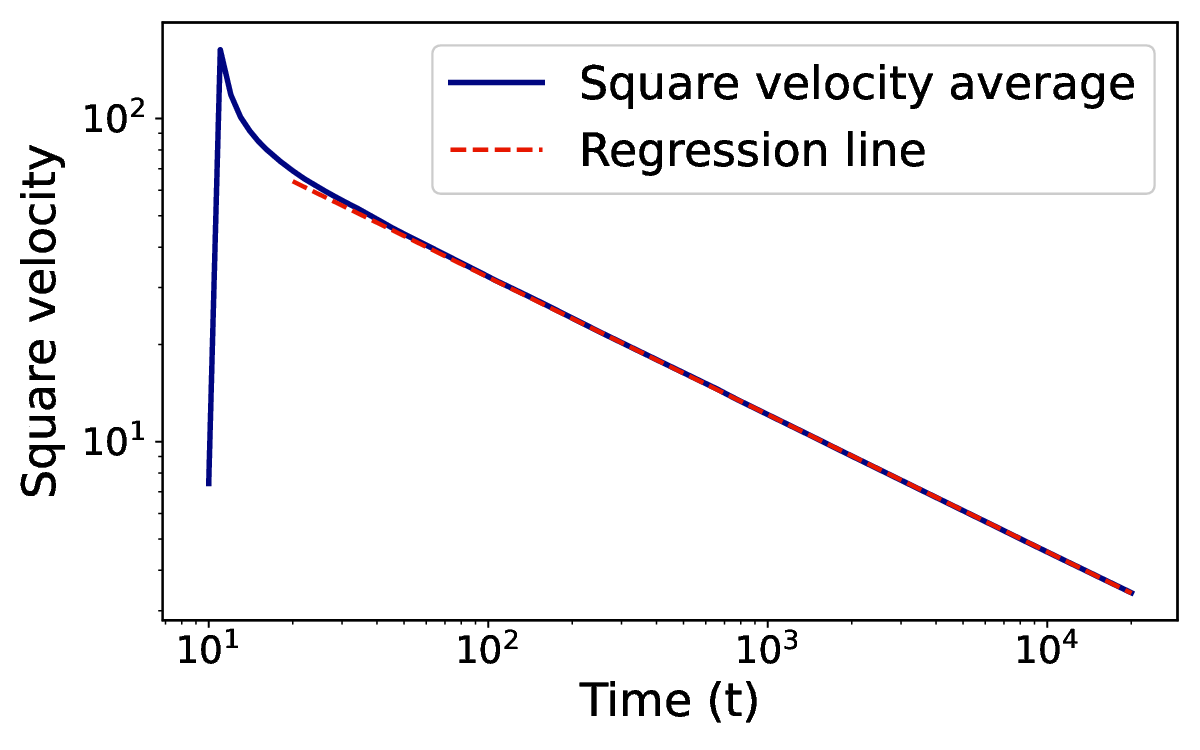}
        \caption{Square velocity time-average. Slope is $-0.425$, fit with $R^2 = 0.998$.}
        \label{graf_noah}
    \end{subfigure}
    \\
    \begin{subfigure}{0.8\textwidth}
        \centering
        \includegraphics[width=0.62\textwidth]{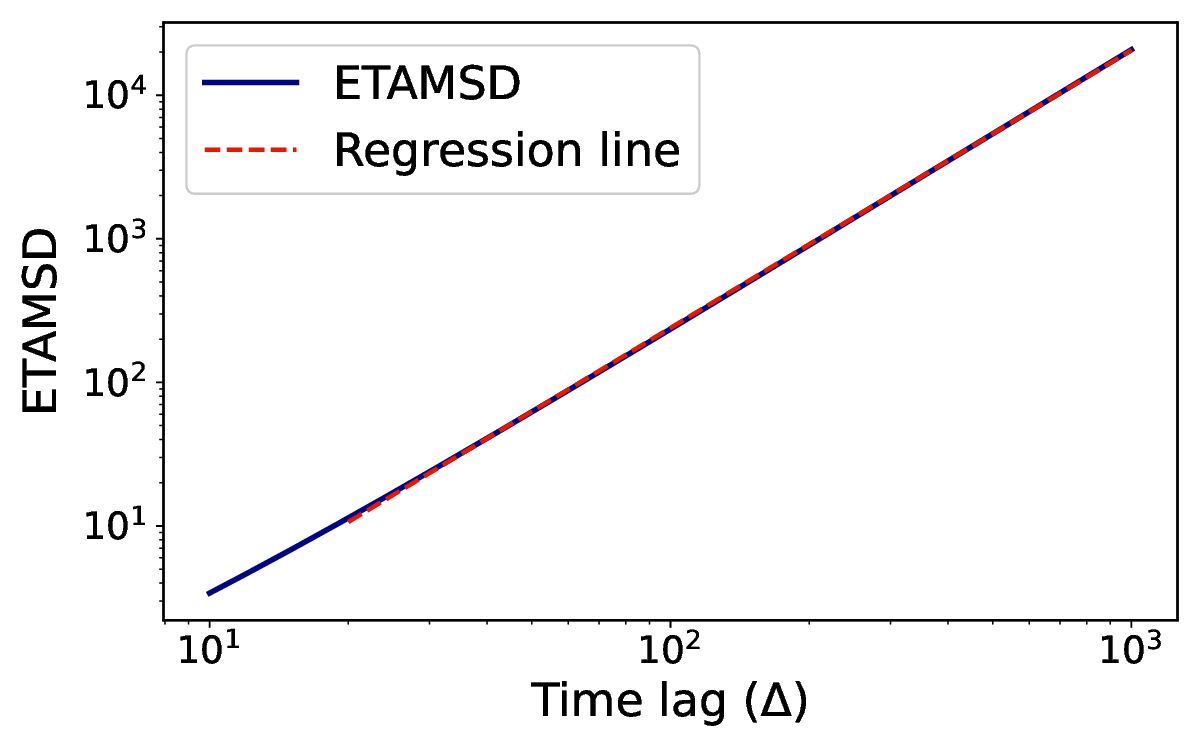}
        \caption{ETAMSD. Slope is $1.934$, fit with $R^2 = 1.000$.}
        \label{graf_joseph}
    \end{subfigure}
    \label{graf_avgs}
\end{figure}

The simulation procedure we followed consists of generating the random interarrival times of the process. The instant of the next arrival is distributed according to \Eref{generalpsi}, therefore the interarrival times can be obtained by generating such random numbers and subtracting the current time $s$ in each case. We input a final time $T$ and the process parameters and obtained a list of arrival times. This list was in turn converted to a list of observations at equally spaced instants according to a sampling period $h$. Repeating this procedure a total of $N$ times yielded an ensemble. Then all averages were computed according to Equations \ref{moses_ts}, \ref{noah_ts}, \ref{etamsd} and \ref{hurst}. The exponents were measured by fitting a line to the corresponding average in logarithmic scale; see as an example Figures \ref{graf_moses}, \ref{graf_noah} and \ref{graf_joseph}, where the absolute velocity, square velocity and ETAMSD plots (for a superdiffusive case) are shown in log-log scale.

\section{Conclusions}\label{sec_conclusiones}

A new anomalous diffusion model based on Generalized Polya Processes was presented. These processes are solutions to the master equation with transition probabilities that depend on two factors: one accounting for state-dependence or contagion, and the other introducing memory-like effects while inducing non-stationarity. We proved that in this model, anomalous diffusion arises from the violation of two conditions of the Central Limit Theorem, as demonstrated by the behaviour of the Moses, Noah and Joseph exponents. The model we introduced accounts for all possible diffusion regimes: subdiffusion, Brownian non-Gaussian, superdiffusion, ballistic and hyperballistic, 
including examples of applications in the latter cases that are not typically considered. The time scaling of the MSD, which entails the self-similarity characteristic of the process, and the heavy-tailed waiting times constitute a universality principle of this type of phenomena. Time scaling plays a crucial role through the Hurst exponent, which in our proposal is directly linked to the model parameters. This scaling and the resulting diffusion regime were shown to be determined from the ratio of two important factors, one that controls the strength of the correlation between displacements and another that governs attenuation with time, acting as two opposing forces. This formulation allows for the construction of models that account for transitions between regimes by properly choosing the relaxation function $\kappa(t)$. Results on this issue will be communicated in future works.

\section*{Acknowledgements}

Two of us NRB and GP wish to thank Universidad Nacional de Tres de Febrero for support under grant no. 80120230100018TF. MFC thanks to PIP-CONICET and CyTUNGS. The authors wish to thank the unknown reviewers for their valuable comments and suggestions that contributed to enhance the article.

\bibliography{bibliography.bib}

\providecommand{\newblock}{}
\begin{thebibliography}{10}
\expandafter\ifx\csname url\endcsname\relax
  \def\url#1{{\tt #1}}\fi
\expandafter\ifx\csname urlprefix\endcsname\relax\def\urlprefix{URL }\fi
\providecommand{\eprint}[2][]{\url{#2}}

\bibitem{Sposini2022}
Sposini V, Krapf D, Marinari E, Sunyer R, Ritort F, Taheri F, Selhuber-Unkel C,
  Benelli R, Weiss M, Metzler R and Oshanin G 2022 {\em Commun. Phys.\/} {\bf
  5} 305 \urlprefix\url{https://doi.org/10.1038/s42005-022-01079-8}

\bibitem{WangBo2012WBdi}
Wang B, Kuo J, Bae S~C and Granick S 2012 {\em Nature Materials\/} {\bf 11}
  481--485 \urlprefix\url{https://doi.org/10.1038/nmat3308}

\bibitem{C4CP02019G}
Jeon J~H, Chechkin A~V and Metzler R 2014 {\em Phys. Chem. Chem. Phys.\/} {\bf
  16}(30) 15811--15817 \urlprefix\url{http://dx.doi.org/10.1039/C4CP02019G}

\bibitem{C4CP03465A}
Metzler R, Jeon J~H, Cherstvy A~G and Barkai E 2014 {\em Phys. Chem. Chem.
  Phys.\/} {\bf 16}(44) 24128--24164
  \urlprefix\url{http://dx.doi.org/10.1039/C4CP03465A}

\bibitem{Kenkre-Montroll-S}
Kenkre V~M, Montroll E~W and Shlesinger M~F 1973 {\em J. Stat. Phys.\/} {\bf 9}
  45--50 \urlprefix\url{https://doi.org/10.1007/BF01016796}

\bibitem{PhysRevB.12.2455}
Scher H and Montroll E~W 1975 {\em Phys. Rev. B\/} {\bf 12}(6) 2455--2477
  \urlprefix\url{https://link.aps.org/doi/10.1103/PhysRevB.12.2455}

\bibitem{ShlesingerDiffusion}
Klafter J, Blumen A and Shlesinger M~F 1987 {\em Phys. Rev. A\/} {\bf 35}(7)
  3081--3085 \urlprefix\url{https://link.aps.org/doi/10.1103/PhysRevA.35.3081}

\bibitem{Metzler_2022_ModellingAnomalous}
Metzler R, Rajyaguru A and Berkowitz B 2022 {\em New J. Phys.\/} {\bf 24}
  123004 \urlprefix\url{https://dx.doi.org/10.1088/1367-2630/aca70c}

\bibitem{PhysRevLett.120.104501}
Albers T and Radons G 2018 {\em Phys. Rev. Lett.\/} {\bf 120}(10) 104501
  \urlprefix\url{https://link.aps.org/doi/10.1103/PhysRevLett.120.104501}

\bibitem{PhysRevE.105.014113}
Albers T and Radons G 2022 {\em Phys. Rev. E\/} {\bf 105}(1) 014113
  \urlprefix\url{https://link.aps.org/doi/10.1103/PhysRevE.105.014113}

\bibitem{Cherstvy_2014_SpaceDependentDiffusivities}
Cherstvy A~G, Chechkin A~V and Metzler R 2014 {\em J. Phys. A: Math. Theor.\/}
  {\bf 47} 485002
  \urlprefix\url{https://dx.doi.org/10.1088/1751-8113/47/48/485002}

\bibitem{Cherstvy_2015_TimeSpaceDependentDiffusivities}
Cherstvy A~G and Metzler R 2015 {\em J. Stat. Mech.\/} {\bf 2015} P05010
  \urlprefix\url{https://dx.doi.org/10.1088/1742-5468/2015/05/P05010}

\bibitem{Cherstvy_2021_TimeDependentDiffusivity}
Cherstvy A~G, Safdari H and Metzler R 2021 {\em J. Phys. D: Appl. Phys.\/} {\bf
  54} 195401 \urlprefix\url{https://dx.doi.org/10.1088/1361-6463/abdff0}

\bibitem{Lemaitre_2023_RandomDiffusivity}
Lemaitre E, Sokolov I~M, Metzler R and Chechkin A~V 2023 {\em New J. Phys.\/}
  {\bf 25} 013010 \urlprefix\url{https://dx.doi.org/10.1088/1367-2630/acb005}

\bibitem{Sposini_2018_RandomDiffusivity}
Sposini V, Chechkin A~V, Seno F, Pagnini G and Metzler R 2018 {\em New J.
  Phys.\/} {\bf 20} 043044
  \urlprefix\url{https://dx.doi.org/10.1088/1367-2630/aab696}

\bibitem{Jain2017_DiffusingDiffusivity}
Jain R and Sebastian K~L 2017 {\em J. Chem. Sci.\/} {\bf 129} 929--937
  \urlprefix\url{https://doi.org/10.1007/s12039-017-1308-0}

\bibitem{Wang_2020_RandomDiffusivity}
Wang W, Cherstvy A~G, Chechkin A~V, Thapa S, Seno F, Liu X and Metzler R 2020
  {\em J. Phys. A: Math. Theor.\/} {\bf 53} 474001
  \urlprefix\url{https://dx.doi.org/10.1088/1751-8121/aba467}

\bibitem{Wang_2020_UnexpectedCrossover}
Wang W, Seno F, Sokolov I~M, Chechkin A~V and Metzler R 2020 {\em New J.
  Phys.\/} {\bf 22} 083041
  \urlprefix\url{https://dx.doi.org/10.1088/1367-2630/aba390}

\bibitem{Palyulin_2019_LevyWalks}
Palyulin V~V, Blackburn G, Lomholt M~A, Watkins N~W, Metzler R, Klages R and
  Chechkin A~V 2019 {\em New J. Phys.\/} {\bf 21} 103028
  \urlprefix\url{https://dx.doi.org/10.1088/1367-2630/ab41bb}

\bibitem{Stella_2010_AnomalousScaling}
Stella A~L and Baldovin F 2010 {\em J. Stat. Mech.\/} {\bf 2010} P02018
  \urlprefix\url{https://dx.doi.org/10.1088/1742-5468/2010/02/P02018}

\bibitem{Tejedor_2010_CorrelatedCTRW_1}
Tejedor V and Metzler R 2010 {\em J. Phys. A: Math. Theor.\/} {\bf 43} 082002
  \urlprefix\url{https://dx.doi.org/10.1088/1751-8113/43/8/082002}

\bibitem{Magdziarz_2012_Correlated_CTRW_2}
Magdziarz M, Metzler R, Szczotka W and Zebrowski P 2012 {\em J. Stat. Mech.\/}
  {\bf 2012} P04010
  \urlprefix\url{https://dx.doi.org/10.1088/1742-5468/2012/04/P04010}

\bibitem{PhysRevE.105.064126}
Akimoto T, Barkai E and Radons G 2022 {\em Phys. Rev. E\/} {\bf 105}(6) 064126
  \urlprefix\url{https://link.aps.org/doi/10.1103/PhysRevE.105.064126}

\bibitem{PolyaLundberg}
Pfeifer D 2004 {Pólya-Lundberg Process} {\em Encyclopedia of Statistical
  Sciences\/} (John Wiley \& Sons, Inc.) ISBN 9780471667193

\bibitem{konno2010exact}
Konno H 2010 {\em Advances in Mathematical Physics\/} {\bf 2010} 504267
  \urlprefix\url{https://onlinelibrary.wiley.com/doi/abs/10.1155/2010/504267}

\bibitem{cha_2014}
Cha J~H 2014 {\em Advances in Applied Probability\/} {\bf 46} 1148--1171
  \urlprefix\url{https://doi.org/10.1239/aap/1418396247}

\bibitem{badia_extensions_gpp}
Badía F~G, Mercier S and Sangüesa C 2019 {\em Methodology and Computing in
  Applied Probability\/} {\bf 21} 1057--1085
  \urlprefix\url{https://hal.archives-ouvertes.fr/hal-01861268}

\bibitem{fendick_whitt_2021}
Fendick K and Whitt W 2021 {\em Journal of Applied Probability\/} {\bf 58}
  484--504 \urlprefix\url{http://doi.org/10.1017/jpr.2020.103}

\bibitem{fendick_whitt_2022}
Fendick K and Whitt W 2022 {\em Queueing Systems\/} {\bf 101} 113--135
  \urlprefix\url{https://doi.org/10.1007/s11134-021-09728-5}

\bibitem{le2015recurrent}
Le~Gat Y 2015 {\em Recurrent Event Modeling Based on the Yule Process:
  Application to Water Network Asset Management\/} (John Wiley \& Sons)

\bibitem{CSFEpidemiologia}
Barraza N~R, Pena G and Moreno V 2020 {\em Chaos, Solitons \& Fractals\/} {\bf
  139} 110297 \urlprefix\url{https://doi.org/10.1016/j.chaos.2020.110297}

\bibitem{BarrazaRPIC2023}
Pena G, Barraza N~R and Gambini J 2023 Network traffic modeled on path
  dependent queues {\em 2023 XX Workshop on Information Processing and Control
  (RPIC)\/} pp 1--6
  \urlprefix\url{https://ieeexplore.ieee.org/document/10530742}

\bibitem{LiGPPReliability}
Li S, Dohi T and Okamura H 2023 {\em IEEE Transactions on Reliability\/} {\bf
  72} 1540--1555

\bibitem{PenaMorenoBarraza2022}
Pena G, Moreno V and Barraza N~R 2022 Stochastic modeling of the mean time
  between software failures: A review {\em System Assurances: Modeling and
  Management\/} Emerging Methodologies and Applications in Modelling,
  Identification and Control ed Johri P, Anand A, Vain J, Singh J and Quasim M
  (Amsterdam, The Netherlands: Elsevier Science) chap~20
  \urlprefix\url{https://www.elsevier.com/books/system-assurances/johri/978-0-323-90240-3}

\bibitem{Pena2022MeasuringCS}
Pena G, Moreno V and Barraza N~R 2022 {\em Epidemiologic Methods\/} {\bf 12}
  20220106 \urlprefix\url{https://doi.org/10.1515/em-2022-0106}

\bibitem{Feller1}
Feller W 1968 {\em An introduction to probability theory and its
  applications\/} 3rd ed vol~1 (Hoboken, NJ, USA: Wiley) ISBN 0471257087

\bibitem{hanggi_convolutionless}
Hänggi P and Talkner P 1978 {\em Physics Letters A\/} {\bf 68} 9--11
  \urlprefix\url{https://www.sciencedirect.com/science/article/pii/0375960178907405}

\bibitem{Monthus_2021}
Monthus C 2021 {\em J. Stat. Mech.\/} {\bf 2021} 063211
  \urlprefix\url{https://dx.doi.org/10.1088/1742-5468/ac06c0}

\bibitem{klugman}
Klugman S~A, Panjer H~H and Willmot G~E 2013 {\em Loss models: Further
  topics\/} Wiley Series in Probability and Statistics (Hoboken, NJ, USA: John
  Wiley \& Sons) ISBN 9781118573747

\bibitem{sendova}
Sendova K~P and Minkova L~D 2019 {\em Stochastics\/} {\bf 92} 814--832
  \urlprefix\url{https://doi.org/10.1080/17442508.2019.1666132}

\bibitem{fendick_whitt_gaussmarkov}
Fendick K and Whitt W 2024 {\em Operations Research Letters\/} {\bf 52} 107062
  \urlprefix\url{https://doi.org/10.1016/j.orl.2023.107062}

\bibitem{app12199965}
Zavyalov A, Zotov O, Guglielmi A and Klain B 2022 {\em Applied Sciences\/} {\bf
  12} 9965 \urlprefix\url{https://www.mdpi.com/2076-3417/12/19/9965}

\bibitem{Guglielmi2016}
Guglielmi A~V 2016 {\em Izvestiya, Physics of the Solid Earth\/} {\bf 52}
  785--786 \urlprefix\url{https://doi.org/10.1134/S1069351316050165}

\bibitem{VollmerAnomalousDiffusion2021}
Vollmer J, Rondoni L, Tayyab M, Giberti C and Mej\'{\i}a-Monasterio C 2021 {\em
  Phys. Rev. Res.\/} {\bf 3}(1)
  \urlprefix\url{https://doi.org/10.1103/PhysRevResearch.3.013067}

\bibitem{bng_diffusion}
Marcone B, Nampoothiri S, Orlandini E, Seno F and Baldovin F 2022 {\em J. Phys.
  A: Math. Theor.\/} {\bf 55} 354003
  \urlprefix\url{https://dx.doi.org/10.1088/1751-8121/ac83fd}

\bibitem{Doerries_2023_EmergentAnomalous}
Doerries T~J, Metzler R and Chechkin A~V 2023 {\em New J. Phys.\/} {\bf 25}
  063009 \urlprefix\url{https://dx.doi.org/10.1088/1367-2630/acd950}

\bibitem{Nampoothiri_2022_BrownianNonGaussian}
Nampoothiri S, Orlandini E, Seno F and Baldovin F 2022 {\em New J. Phys.\/}
  {\bf 24} 023003 \urlprefix\url{https://dx.doi.org/10.1088/1367-2630/ac4924}

\bibitem{PerriAmatoZimbardo}
Perri S, Amato E and Zimbardo G 2016 {\em Astronomy \& Astrophysics\/} {\bf
  596} A34 \urlprefix\url{https://doi.org/10.1051/0004-6361/201628767}

\bibitem{Zimbardo}
Zimbardo G, Amato E, Bovet A, Effenberger F, Fasoli A, Fichtner H, Furno I,
  Gustafson K, Ricci P and Perri S 2015 {\em Journal of Plasma Physics\/} {\bf
  81} 495810601 \urlprefix\url{https://doi.org/10.1017/S0022377815001117}

\bibitem{Hanggi}
Siegle P, Goychuk I and H\"anggi P 2010 {\em Phys. Rev. Lett.\/} {\bf 105}(10)
  100602
  \urlprefix\url{https://link.aps.org/doi/10.1103/PhysRevLett.105.100602}

\bibitem{HyperballisticTransport}
Gamba D, Cui B and Zaccone A 2024 {\em Phys. Rev. E\/} {\bf 110}(5) 054137
  \urlprefix\url{https://link.aps.org/doi/10.1103/PhysRevE.110.054137}

\bibitem{VilkAnomalousDiffusion2022}
Vilk O, Aghion E, Avgar T, Beta C, Nagel O, Sabri A, Sarfati R, Schwartz D~K,
  Weiss M, Krapf D, Nathan R, Metzler R and Assaf M 2022 {\em Phys. Rev.
  Res.\/} {\bf 4}(3) 033055
  \urlprefix\url{https://doi.org/10.1103/PhysRevResearch.4.033055}

\bibitem{AghionMosesNoahJoseph2021}
Aghion E, Meyer P~G, Adlakha V, Kantz H and Bassler K~E 2021 {\em New J.
  Phys.\/} {\bf 23} 023002
  \urlprefix\url{https://dx.doi.org/10.1088/1367-2630/abd43c}

\bibitem{chen_anomalous}
Chen L, Bassler K~E, McCauley J~L and Gunaratne G~H 2017 {\em Phys. Rev. E\/}
  {\bf 95}(4) 042141 \urlprefix\url{https://doi.org/10.1103/PhysRevE.95.042141}

\bibitem{vitali_anomalous_diffusion}
Vitali S, Paradisi P and Pagnini G 2022 {\em J. Phys. A: Math. Theor.\/} {\bf
  55} 224012 \urlprefix\url{https://dx.doi.org/10.1088/1751-8121/ac677f}

\bibitem{penagithub-nhbpgen}
Pena G 2024 {NHBPGen} {GPP simulation source code, implemented in Python.}
  \urlprefix\url{https://doi.org/10.5281/zenodo.13947182}

\end{thebibliography}

\end{document}


\begin{center}
\LARGE \bf {A non-homogeneous, non-stationary and path-dependent Markov anomalous diffusion model. Supplementary Material}
\end{center}

We present here some relevant calculations regarding the 3p-BPM process as an anomalous diffusion model.

\section{Analytical calculation of the exponents}\label{ap_moses}

\subsection{Moses}

The Moses exponent is defined as:
\begin{equation}\label{moses_ts}
    \left\langle \overline{|v(t)|} \right\rangle = 
 \left\langle \frac{\Delta}{t}\sum_{j=1}^{t/\Delta} \frac{|\delta X_{j,\Delta}|} {\Delta} \right\rangle \propto t^{M - 1/2}.
\end{equation}

The mean value function of the increments is given by:
\begin{equation}
    \label{mvincre}
    E[\delta(s,t)] = \frac{\beta}{\gamma} \left[ (1 + \rho t)^\frac{\gamma}{\rho} - (1 + \rho s)^\frac{\gamma}{\rho} \right]. 
\end{equation}

Substituting Equation \ref{mvincre} into Equation \ref{moses_ts} the ensemble average of $\delta_{j,\Delta}$ results:

\begin{equation*}
    E[\delta X_{j,\Delta}] = \frac{\beta}{\gamma} \;  \left\lbrace (1 + \rho j \Delta)^\frac{\gamma}{\rho} - [1 + \rho (j - 1)\Delta]^\frac{\gamma}{\rho} \right\rbrace.
\end{equation*}
Thus
\begin{equation*}
    \frac{\Delta}{t}\sum_{j = 1}^{t/\Delta} \frac{E[\delta X_{j,\Delta}]}{\Delta} \propto \frac{1}{t}\sum_{j = 1}^{t/\Delta} \left\lbrace (1 + \rho j \Delta)^\frac{\gamma}{\rho} - [1 + \rho (j - 1)\Delta]^\frac{\gamma}{\rho} \right\rbrace.
\end{equation*}
This is a finite telescopic sum, which yields:
\begin{equation}
\label{moses-2}
    \frac{\Delta}{t}\sum_{j = 1}^{t/\Delta} \frac{E[\delta X_{j,\Delta}]}{\Delta} \propto \frac{1}{t} (1 + \rho t)^\frac{\gamma}{\rho} - 1 \propto t^{\frac{\gamma}{\rho} - 1}.
\end{equation}
From Equations \ref{moses_ts} and \ref{moses-2} results $\frac{\gamma}{\rho} = M + \frac{1}{2}$, which is exactly what was observed in the simulations.

\subsection{Noah}\label{ap_noah}

The Noah exponent is defined as:

\begin{equation}\label{noah_ts}
    \left\langle \overline{v^2(t)} \right\rangle = \left\langle \frac{\Delta}{t}\sum_{j=1}^{t/\Delta} \left( \frac{\delta X_{j,\Delta}}{\Delta} \right)^2 \right\rangle \propto t^{2L + 2M - 2}.
\end{equation}

Since the increments of a GPP are negative binomial distributed, they always have finite variance. Hence the Noah exponent must be $1/2$. We derive this result explicitly by computing Equation \ref{noah_ts}. We begin by writing the variance of the increment $\delta(s,t)$ of a 3p-BPM process:
\begin{equation*}
    Var(\delta(s,t)) = \frac{\beta}{\gamma} \left[ \left((1 + \rho t)^\frac{\gamma}{\rho} - (1 + \rho s)^\frac{\gamma}{\rho}\right)^2 + (1 + \rho t)^\frac{\gamma}{\rho} - (1 + \rho s)^\frac{\gamma}{\rho} \right].
\end{equation*}
Thus
\begin{equation*}
    E[\delta^2(s,t)] = Var(\delta(s,t) + (E[\delta(s,t)])^2 \propto a^2(s,t) + a(s,t), 
\end{equation*}
where $a(s,t) = (1 + \rho t)^\frac{\gamma}{\rho} - (1 + \rho s)^\frac{\gamma}{\rho}$.
Now recall that we are interested in computing increments over intervals of equal length $\Delta$. Thus, we define
\begin{equation*}
    a_j \equiv a(\Delta(j-1), \Delta j) = (1 + \rho \Delta j)^\frac{\gamma}{\rho} - (1 + \rho \Delta(j-1))^\frac{\gamma}{\rho},
\end{equation*}
for all $j\in\mathbb{N}$.

In what follows we will make extensive use of the fact that in Equation \ref{noah_ts}, $t \gg \Delta$. Let us write
\begin{equation*}
    \frac{\Delta}{t}\sum_{j=1}^{t/\Delta} E\left[ \left( \frac{\delta X_{j,\Delta}}{\Delta} \right)^2 \right] \propto \frac{1}{t} \sum_{j=1}^{t/\Delta} a^2_j + \frac{1}{t}\sum_{j=1}^{t/\Delta} a_j.
\end{equation*}

The second summation is the same one that we calculated in \ref{ap_moses}, and we know it to be $\propto t^{\frac{\gamma}{\rho} - 1}$. Thus we need only to calculate the first summation. Since $a_j$ has the form of a finite difference and $\Delta \ll t$, we can approximate it by a derivative:
\begin{equation*}
    a_j \approx \Delta \left.\frac{d}{dx}(1 + \rho\Delta x)^\frac{\gamma}{\rho}\right|_{x=j-1} = \gamma \Delta^2 (1 + \rho\Delta(j-1))^{\frac{\gamma}{\rho} - 1}.
\end{equation*}
The summation then becomes:
\begin{equation*}
    \frac{1}{t}\sum_{j=1}^{t/\Delta} a^2_j \approx \frac{1}{t}\sum_{j=1}^{t/\Delta} \gamma \Delta^2 (1 + \rho\Delta(j-1))^{\frac{\gamma}{\rho} - 1} \propto \frac{1}{t}\sum_{j=1}^{t/\Delta} j^{2\frac{\gamma}{\rho} - 2}.
\end{equation*}
But again, since the time step $\Delta \ll t$, this summation can be approximated by the definite integral: 
\begin{equation*}
    \frac{1}{t}\sum_{j=1}^{t/\Delta} j^{2\frac{\gamma}{\rho} - 2} \approx \frac{1}{t} \int_1^{t/\Delta} x^{2\frac{\gamma}{\rho} - 2} dx \propto t^{2\frac{\gamma}{\rho} - 2}.
\end{equation*}
Therefore we have
\begin{equation*}
    \frac{\Delta}{t}\sum_{j=1}^{t/\Delta} E\left[ \left( \frac{\delta X_{j,\Delta}}{\Delta} \right)^2 \right] \propto t^{2\frac{\gamma}{\rho} - 2} + t^{\frac{\gamma}{\rho} - 1} = \tau^2 + \tau,
\end{equation*}
where $\tau = t^{\frac{\gamma}{\rho} - 1}$. For any fixed parameters and sufficiently large $t$ the linear term can be neglected, so the entire expression should scale as $t^{2\frac{\gamma}{\rho} - 2}$.

Finally, note that since $M = \frac{\gamma}{\rho} - 1/2$, Equation \ref{noah_ts} must scale as $t^{2L + 2\frac{\gamma}{\rho} - 3}$. Therefore we have $L=1/2$ for any set of the parameters. This is consistent with the variance of the increments being always finite.

\section{Joseph}\label{ap_joseph}

We appeal to Eq. 9 of \cite{AghionMosesNoahJoseph2021} and compute the autocorrelation function of the increments. To simplify notation, we assume without loss of generality that increments are sampled over intervals of unity length, as suggested in \cite{chen_anomalous}. Then, the autocorrelation function becomes:
\begin{equation*}
    \frac{E[\delta(t-1, t)\delta(t-1+\Delta, t+\Delta)] }{ E[\delta^2(t-1, t)]} \propto \Delta^{2J - 2}.
\end{equation*}

Since the denominator does not depend on $\Delta$, we need only to calculate the joint moment of the increments. The joint distribution of two disjoint increments is given by
\begin{equation}\label{eq_multinomial}
   P(\delta(s_1, t_1) = n_1, \delta(s_2, t_2) = n_2) = \frac{\Gamma\left(\frac{\beta}{\gamma} + n_1 + n_2\right)}{\Gamma\left(\frac{\beta}{\gamma}\right) n_1! n_2!} p_0^{\frac{\beta}{\gamma}}  p_1^{n_1}  p_2^{n_2},
\end{equation}
where
\begin{eqnarray*}
	 p_0 = \frac{e^{-\gamma K(0, s_1)} e^{-\gamma K(s_1,t_1)} e^{-\gamma K(s_2, t_2)}}{1 - \left(1 - e^{-\gamma K(0, s_1)}\right)e^{-\gamma K(s_1, t_1)}e^{-\gamma K(s_2, t_2)}},\\
	 p_1 = \frac{\left(1 - e^{-\gamma K(s_1, t_1)}\right) e^{-\gamma K(s_2, t_2)}}{1 - \left(1 - e^{-\gamma K(0, s_1)}\right)e^{-\gamma K(s_1, t_1)}e^{-\gamma K(s_2, t_2)}}, \\
	 p_2 = \frac{1 - e^{-\gamma K(s_2, t_2)}}{1 - \left(1 - e^{-\gamma K(0, s_1)}\right)e^{-\gamma K(s_1, t_1)}e^{-\gamma K(s_2, t_2)}} .
\end{eqnarray*}
Since $p_0 + p_1 + p_2 = 1$, this is a negative multinomial distribution. Equation \ref{eq_multinomial} can be derived by considering the known formula for $P(\delta(s_1, t_1) = n_1, \delta(s_2, t_2)=n_2, N(s_1) = k$ (see \cite{cha_2014}) and summing over all possible $k$. Plugging the expression for $K(s,t)$ in our model yields:
\begin{eqnarray*}
    p_0 = \frac{ \left( \frac{1}{1 + \rho s_1} \right)^{\frac{\gamma}{\rho}} \left( \frac{1 + \rho s_1}{1 + \rho t_1} \right)^\frac{\gamma}{\rho} \left( \frac{1 + \rho s_2}{1 + \rho t_2} \right)^\frac{\gamma}{\rho} }{L(s_1,t_1, s_2, t_2)}, \\
    p_1 = \frac{ \left[1 - \left( \frac{1 + \rho s_1}{1 + \rho t_1} \right)^\frac{\gamma}{\rho} \right] \left( \frac{1 + \rho s_2}{1 + \rho t_2} \right)^\frac{\gamma}{\rho} }{L(s_1,t_1, s_2, t_2)}, \\
    p_2 = \frac{ 1 - \left( \frac{1 + \rho s_2}{1 + \rho t_2} \right)^\frac{\gamma}{\rho} }{L(s_1,t_1, s_2, t_2)}, \\
    L(s_1,t_1, s_2, t_2) = 1 - \left[ 1 - \left( \frac{1}{1 + \rho s_1} \right)^\frac{\gamma}{\rho} \right] \left( \frac{1 + \rho s_1}{1 + \rho t_1} \right)^\frac{\gamma}{\rho} \left( \frac{1 + \rho s_2}{1 + \rho t_2} \right)^\frac{\gamma}{\rho}.
\end{eqnarray*}

Taking $s_1 = t-1$, $t_1 = t$, $s_2 = t + \Delta - 1$, $s_2 = t + \Delta$ we see that for large $\Delta$ all $p_0$, $p_1$ and $p_2$ scale as constants. Hence:
\begin{equation*}
     Cov(\delta(t-1, t)\delta(t - 1 + \Delta, t + \Delta)) = r\frac{p_1p_2}{p_0} \propto \Delta^0.
\end{equation*}

Now we need to compute $E[\delta(s_1, t_1)]$ and $E[\delta(s_2, t_2)]$. It is known that their marginal distributions are negative binomial with suitable scaled parameters (consult our main paper or \cite{cha_2014}). The distribution of the last increment $\delta(s_2, t_2)$ is negative binomial with $\tilde{p} = \frac{p_0}{1 - p_1}$. Thus:
\begin{eqnarray*}
    E[\delta(t - 1, t)] = r \frac{p_1(1 - p_2)}{p_0(1 - p_2)} = r \frac{p_1}{p_0} \propto \Delta^0, \\
    E[\delta(t + \Delta - 1, t + \Delta)] = r \frac{p_2(1 - p_1)}{p_0(1 - p_1)} = r \frac{p_2}{p_0} \propto \Delta^0.
\end{eqnarray*}

Since both the covariance and the marginal expectations scale as constants, the joint moment does. Therefore $2J - 2 = 0$, which implies $J=1$. For a 3p-BPM model, the Joseph exponent is independent of the model parameters and is always equal to 1, meaning that the increments are maximally correlated. 

\section{Asymptotic limit of the autocorrelation function}

Substituting the value of $K(t)$ for the 3p-BPM process in the autocovariance and variance formulae, we get:
\begin{eqnarray*}
    Cov[X(s),X(t)] = (1 + \rho t)^\frac{\gamma}{\rho}  \left[ (1 + \rho s)^\frac{\gamma}{\rho} - 1 \right] \\
    Var[X(t)] =  (1 + \rho t)^\frac{\gamma}{\rho} \left[ (1 + \rho t)^\frac{\gamma}{\rho} - 1 \right]
\end{eqnarray*}

Thus the autocorrelation function is:
\begin{equation*}
    R[X(s),X(t)] = \sqrt{\frac{(1 + \rho s)^\frac{\gamma}{\rho} - 1}{(1 + \rho s)^\frac{\gamma}{\rho}}} \cdot \sqrt{\frac{(1 + \rho t)^\frac{\gamma}{\rho}}{ (1 + \rho t)^\frac{\gamma}{\rho} - 1 }} \; \underset{t\to\infty}{\longrightarrow} \; \sqrt{1 - \frac{1}{(1 + \rho s)^\frac{\gamma}{\rho}}}
\end{equation*}

This shows that the autocorrelation does not decrease to zero, but tends to a constant value that in turn depends on $\gamma/\rho$. This limit approaches one for large $\gamma/\rho$, and approaches zero when $\gamma/\rho$ is close to zero. 

\section{Parameters used for the simulation}

Table \ref{tab_param_simulaciones} describes all relevant parameters used for the simulations and the averaging procedures. $N$ is the number of simulations, $T$ is the total observed time, $\Delta$ is the time gap across which the increments are computed and $h$ is the sampling period. To observe the scaling, we computed the averages from small values and up to $T$ and $\Delta$ respectively. The complete code is available at \cite{penagithub-nhbpgen}.

\begin{table}[!ht]
    \renewcommand{\arraystretch}{1.5}
    \caption{Parameters used for the simulations and averaging.}
    \label{tab_param_simulaciones}
    \centering\vspace{1em}
    \begin{tabular}{|c|c|c|c|c|}
        \hline
        $\gamma/\rho$ & $N$ & $T$ & $\Delta$ & $h$ \\
        \hline\hline
        $1/4$ & $10^3$ & $10^6$ & $10^4$ & $1$ \\
        \hline
        $1/2$ & $10^3$ & $10^5$ & $10^3$ & $1$ \\
        \hline
        $3/4$ & $10^3$ & $2 \cdot 10^4$ & $10^3$ & $1$ \\ 
        \hline
        $1$ & $10^3$ & $8 \cdot 10^3$ & $200$ & $1$ \\ 
        \hline
        $5/4$ & $10^3$ & $2 \cdot 10^3$ & $100$ & $1$ \\
        \hline
        $3/2$ & $10^3$ & $10^3$ & $100$ & $1$ 
        \\
        \hline
        $2$ & $10^3$ & $200$ & $50$ & $1$ 
        \\
        \hline
    \end{tabular}
\end{table}

\bibliography{biblio_app.bib}